\documentclass[%
preprint,
 amsmath,amssymb,
 aps,
]{revtex4-1}

\usepackage{graphicx}% Include figure files
\usepackage{bm}% bold math
\usepackage{hyperref}% add hypertext capabilities
\hypersetup{
    colorlinks=true,
    linkcolor=red,
    filecolor=magenta,      
    urlcolor=blue, %cyan,
    pdftitle={Sharelatex Example},
    bookmarks=true,
    pdfpagemode=FullScreen,
}

\begin{document}

%\preprint{APS/123-QED}

\title{%Classical collective Behaviours Emerging from Quantum Mechanics:\\
Quantum Central Limit Theorems, Emergence of Classicality\\and Time-dependent Differential Entropy}% Force line breaks with \\

\author{Tien D. Kieu}
 \email{tien.d.kieu@gmail.com}
\affiliation{%
Centre for Quantum Technology Theory,\\
Swinburne University of Technology, Victoria, Australia
}%

% \collaboration{MUSO Collaboration}%\noaffiliation

%  \homepage{http://www.Second.institution.edu/~Charlie.Author}
% \affiliation{
%  Second institution and/or address\\
%  This line break forced% with \\
% }%
% \affiliation{
%  Third institution, the second for Charlie Author
% \affiliation{%
%  Authors' institution and/or address\\
%  This line break forced with \textbackslash\textbackslash
% }%

% \collaboration{CLEO Collaboration}%\noaffiliation

%\date{}% It is always \today, today,
\date{\today}
\begin{abstract}
We derive some Quantum Central Limit Theorems for expectation values of macroscopically coarse-grained observables, which are functions of coarse-grained hermitean operators. Thanks to the hermicity constraints, we obtain positive-definite distribution for the expectation values of observables. These probability distributions open some pathway for an emergence of classical behaviours in the limit of infinitely large number of identical and non-interacting quantum constituents. This is in contradistinction to other mechanisms of classicality emergence due to environmental decoherence and consistent histories. The probability distributions so derived also enable us to evaluate the nontrivial {\it time-dependence} of certain differential entropies.

% \begin{description}
% Secondary publications and information retrieval purposes.
% \item[PACS numbers]
% May be entered using the \verb+\pacs{#1}+ command.
% You may use the \texttt{description} environment to structure your abstract;
% use the optional argument of the \verb+\item+ command to give the category of each item. 
% \end{description}
\end{abstract}

\pacs{Valid PACS appear here}% PACS, the Physics and Astronomy
\maketitle
\section{Opening remarks}
The Central Limit Theorem (CLT)~\cite{Fischer} for sums of independent identically distributed (iid) random variables is one of the most fundamental pillars of classical probability theory. It and various generalisations~\cite{Feller} have found numerous applications in diverse fields including mathematics, physics, information theory, economics, finance and psychology.

The CLT has also been generalised to various quantum versions~\cite{Cushen, Hepp, Hepp2, Giri, Quaegebeur, Goderis, Lenczewski, Dorlas, Jaksic} which have also found many applications in quantum statistical mechanics, quantum field theory, quantum information theory, graph theory, non-commutative algebras and non-commutative stochastic processes.

In this paper we derive a quantum version of the CLT for expectation values of hermitian operators only, and not of general operators. This hermicity constraint for observables results in positive-definite probability distributions -- in contradistinction to the Wigner function, which is a quasiprobability distribution and although is real-valued but not positive-definite in general. Our probability distributions below,~(\ref{gaussian}) and~(\ref{QCLT}), are also {\it unique} and {\it independent} of the operators being considered. We employ in our derivation a renormalisation blocking approach to obtain {\it explicit} expressions for the probability distributions. These are presented in the next three sections.

Note that, on the other hand, previous versions of quantum CLT consider general operators, including non-hermitian ones, and thus do not explicitly express the resulting quasi-disributions but only {\it implicitly} through expectation values with gaussian states. The exact forms of those quasi-distributions, as such, may also be dependent on the operators being considered.

With the explicit forms of our so-derived probability distributions, there affords a pathway for emergence of classical behaviours from quantum mechanics of a system having non-interacting constituents when the number of constituents is taken to infinity. This is discussed in section~\ref{sec5}. Our pathway for an emergence of classicality is quite different from that afforded by decoherence and/or consistent histories. 

We then use our explicit probability distributions for the evaluation of particular form of differential entropy for some simple quantum systems in section~\ref{sec6}. In the literature for both open and closed quantum systems, different information-theoretical entropy measures have been discussed~\cite{Zurek, Omnes, Anastopoulos, Leipnik, Dodonov}. The measure of differential entropy we employ is a special case of relative entropy, argued for based on the considerations by Jaynes~\cite{Jaynes}.

The last section of the paper contains our concluding remarks.

\section{Quantum central limit theorem and heuristic renormalisation blocking}

Renormalisation group blocking plays a central role in understanding emerging bulk behaviours and collective phenomena. Heuristically, one could start with a path integral or partition function in some set of fundamental microscopic variables / operators. As an illustration, let us take the path integral expression for  a quantum system having the action ${\cal S}[\xi]$ in the fundamental field variable $\xi$,
\begin{eqnarray}
{\cal Z} &=& \int \exp\{i{\cal S}[\xi]\} {\cal D}\xi.
\end{eqnarray}
In deriving coarse-graining behaviours from the system,
we introduce the coarse-grained field variable $\Xi$ as a function of the fundamental variables $\xi$ in some chosen blocking scheme $\Xi_j = h(\xi)$, as in an averaging scheme, for example, 
\begin{eqnarray}
\Xi_j &=& \frac{1}{V_j}\sum_{\xi_i \in {\rm block}\; j}\xi_i,
\end{eqnarray}
where $V_j$ is a measure of the ``volume" of each block.
We have to choose the blocking function $h$ in such a way that the coarse-grained variables are not growing indefinitely in magnitude when we keep on coarse-graining the coarse-grained variables successively to the next level -- hence the volume denominator in our example above.

The expectation value of a quantum operator $\langle F(\Xi)\rangle$ of the coarse-grained variables could then be expressed as
\begin{eqnarray}
\langle F(\Xi(\xi))\rangle &=& \frac{1}{{\cal Z}}\int F(\Xi(\xi)) \exp\{i{\cal S}[\xi]\} {\cal D}\xi.\label{exptn}
\end{eqnarray}
To convert the last path integral in ${\cal D}\xi$ to that in ${\cal D}\Xi$, we insert the resolution of unity
\begin{eqnarray}
1 &=& \int \delta(\Xi - h(\xi)){\cal D}\Xi
\end{eqnarray}
into (\ref{exptn}) then interchange the order of integration to obtain
\begin{eqnarray}
\langle F(\Xi)\rangle &=& \frac{1}{{\cal Z}}\int F(\Xi) \exp\{i{\cal S}'[\Xi]\} {\cal D}\Xi,
\end{eqnarray}
where
\begin{eqnarray}
\exp\{i{\cal S}'[\Xi]\} &=& \int \delta(\Xi - h(\xi))
\exp\{i{\cal S}[\xi]\}{\cal D}\xi.
\end{eqnarray}

Successive repeating of the last expression defines a renormalisation group flow.

We will phrase the quantum central limit theorems in this paper as a restricted renormalisation blocking in the sense that we consider only the expectation values of hermitean operators and not the full path integral / partition function for arbitrary operators.

Our restricted consideration results in positive-definite measures which can be interpreted as probability measures, from which the fixed-point distributions of the renormalisation blocking emerge.

\section{Central Limit Theorem for single hermitian variable}
The centre of mass, or intensive variables in general, of a composite systems of $N$ components can be expressed as
\begin{eqnarray}
\hat X &=& \frac{1}{N}\left(\hat x_1\otimes \hat 1 _2 \cdots \otimes \hat 1_N  + \hat 1_1\otimes \hat x _2 \cdots \otimes \hat 1_N + \cdots + \hat 1_1\otimes \hat 1 _2 \cdots \otimes \hat x_N\right) \nonumber \\
&\equiv& \frac{1}{N}\sum_{i=1}^{N} \hat x_i
\end{eqnarray}
We now consider a system with identical and non-interacting components (such as the case of an ideal gas)
\[|\Phi\rangle = \otimes_i^N |\phi_i\rangle,\]
where $|\phi_i\rangle = |\phi\rangle$, for all $i$.

With some general function $f$, we obtain the following result for $N\gg1$
\begin{eqnarray}
\left\langle\Phi \left| f\left(\frac{1}{N}\sum_i^N \hat x_i\right)\right|\Phi \right\rangle 
&\stackrel{N\to \infty}{\longrightarrow}& \frac{1}{(\sigma_x/\sqrt N)2\pi \sqrt{2\pi}}\int dX f(X) \exp{\left\{-\frac{(X - \langle{x}\rangle)^2}{2(\sigma_x/\sqrt N)^2}\right\}}, \label{gaussian}
\end{eqnarray}
where 
\begin{eqnarray}
\langle{x}\rangle &\equiv& \int x |\langle x|\phi \rangle|^2  = \langle \phi|\hat x|\phi\rangle,\\
\sigma_x^2 &\equiv& \langle{x^2}\rangle - \langle{x}\rangle^2.
\end{eqnarray}
The derivation of the above is given in Appendix~\ref{app2}.

In particular, we can derive, as a special case from the above, the probability density for finding $X$ around some $X_0$,
\begin{eqnarray}
\left\langle\Phi \left|\delta\left( \hat X - X_0\hat {\bf 1}\right)\right|\Phi\right\rangle &\sim& \int \delta\left( X - X_0\right)
\exp{\left\{-\frac{(X - \langle{x}\rangle)^2}{2(\sigma_x/\sqrt N)^2}\right\}}d X\nonumber\\
&\sim& \exp{\left\{-{(X_0 - \langle{x}\rangle)^2}/{2(\sigma_x/\sqrt N)^2}\right\}},
\end{eqnarray}
which is a gaussian distribution.

We could estimate from the derivation of the above that the size of the system should satisfy the condition $N \gg \left|\langle x^3\rangle/\langle x^2\rangle\right|$ for the approximation.

We can also easily generalise the result to the case when the initial state is a mixed state instead of being pure.

Note also that the above result can be readily generalised  to the case when
\[ f\left(\frac{1}{N}\sum_i^N x_i\right) \longrightarrow f\left(\frac{1}{N^m}\sum_i^{N} g(x_i)\right),\; m\ge 1 \]
where $m$ is integer and $g()$ is some arbitrary function.

In the limit of $N \to \infty$ the gaussian distribution in~(\ref{gaussian}) converges to a delta distribution,
\begin{eqnarray}
\frac{1}{(\sigma_x/\sqrt N)2\pi \sqrt{2\pi}} \exp{\left\{-\frac{(X - \langle{x}\rangle)^2}{2(\sigma_x/\sqrt N)^2}\right\}} &\stackrel{N\to \infty}{\longrightarrow}& \delta(X - \langle{x}\rangle).
\end{eqnarray}
We thus have from~(\ref{gaussian}), for arbitrarily finite integer $m$,
\begin{eqnarray}
\left\langle\Phi \left| {\hat X}^m\right|\Phi \right\rangle 
&\stackrel{N\to \infty}{\longrightarrow}& \int dX  \;X^m\delta(X - \langle{x}\rangle) = \langle x\rangle^m.
\end{eqnarray}
This is an indication of an emergence of classical behaviours for macroscopically blocked variable $X$, as the right hand side of the last expression contains $\langle x\rangle^m$ rather than $\langle x^m\rangle$. 

In order to verify such emergence we will need to further consider quantum mechanically non-commuting variables in the next section.

\section{A Central Limit Theorem for non-commuting variables}
We additionally consider the momentum operators $\hat{p}_i$, the non-commuting conjugate variables of the position operators. and introduce the blocked variable $\hat{P}$
\begin{eqnarray}
\hat P &=& \frac{1}{N}\left(\hat p_1\otimes \hat 1 _2 \cdots \otimes \hat 1_N  + \hat 1_1\otimes \hat p _2 \cdots \otimes \hat 1_N + \cdots + \hat 1_1\otimes \hat 1 _2 \cdots \otimes \hat p_N\right) \nonumber \\
&\equiv& \frac{1}{N}\sum_{i=1}^{N} \hat p_i
\end{eqnarray}

While $\hat X$ of the last section is the centre of mass, this blocked variable $\hat P$ corresponds to a measure of the velocity of the centre of mass.

Even for system of interacting components, we have
\begin{eqnarray}
\left[ \hat X, \hat P\right] &=& \frac{1}{N^2}\left[ \sum_{i}^N \hat x_i, \sum_{j}^N \hat p_j\right] \nonumber\\
&=& \frac{1}{N^2}\sum_i^N \left[ \hat x_i, \hat p_i\right] \nonumber\\
&=& {i\hbar}/{N} \nonumber \\
&\stackrel{N\to\infty}{\longrightarrow}& 0
%\label{classicality}
\end{eqnarray}

With
\[\Delta A^2\,\Delta B^2 \ge \left|\frac{1}{2}\langle \{\hat A,\hat B\} \rangle-\langle \hat A\rangle\langle \hat B\rangle\right|^2 + \left|\frac{1}{2i}\langle[\hat A,\hat B]\rangle\right|^2,\]
we have, because of the approximate commutativity above,
\begin{eqnarray}
\Delta X\,\Delta P &\sim& O({1}/{N}) \stackrel{N\to\infty}{\longrightarrow}0. \label{classicality2}
\end{eqnarray}

We now consider a {\it hermitian} combination of 
some finite sum of products of $\hat X$ and $\hat P$, which can be expressed in general as, by constraint of hermiticity,
\begin{eqnarray}
c_{mn}(\hat X)^m (\hat P)^n + c^*_{mn}(\hat P)^n (\hat X)^m,
\end{eqnarray}
where $c_{mn}$ are c-numbers. For expectation values of general observables, we can indeed further restrict the above to real values of $c_{mn}$.

For $N\gg1$, we obtain the following result, of which the derivation is presented in Appendix~\ref{app1},
\begin{eqnarray}
\left\langle\Phi\left|
\sum_{mn} \left(c_{mn} X^m P^n + c^*_{mn} P^n X^m\right)
\right|\Phi\right\rangle  &=& \int dX dP \;\left(2\sum_{mn}\Re(c_{mn})X^m P^n\right){\cal P}_{re}(X,P)\nonumber\\
&& + \int dX dP \;\left(\sum_{mn}\Im(c_{mn})X^m P^n\right){\cal P}_{im}(X,P),\label{QCLT}\nonumber\\
\end{eqnarray}
where the probability distribution for the real parts $\Re(c_{mn})$ is
\begin{eqnarray}
{\cal P}_{re}(X,P) &\sim& 
\exp{\left\{-\frac{\left[(X - \langle{x}\rangle)\cos\theta_+ + (P - \langle{p}\rangle)\sin\theta_+\right]^2}{(\sigma_x^2 + \sigma_p^2+\Delta_+)/N}\right\}}\nonumber\\
&&\times\exp{\left\{-\frac{\left[(P - \langle{p}\rangle)\cos\theta_+ - (X - \langle{x}\rangle)\sin\theta_+\right]^2}{(\sigma_x^2 + \sigma_p^2-\Delta_+)/N}\right\}}, \label{reDist}
\end{eqnarray}
while the probability distribution for the imaginary parts $\Im(c_{mn})$ is
\begin{eqnarray}
{\cal P}_{im}(X,P) &=& 
{\cal N}_1\exp{\left\{-\frac{\left[(X - \langle{x}\rangle)\cos\theta_- + (P - \langle{p}\rangle)\sin\theta_-\right]^2}{(\sigma_x^2 + \sigma_p^2+\Delta_-)/N}\right\}}\nonumber\\
&&\;\;\;\;\;\times\exp{\left\{-\frac{\left[(P - \langle{p}\rangle)\cos\theta_- - (X - \langle{x}\rangle)\sin\theta_-\right]^2}{(\sigma_x^2 + \sigma_p^2-\Delta_-)/N}\right\}}\nonumber\\
&& -{\cal N}_2\exp{\left\{-\frac{(P - \langle{p}\rangle)^2}{2\sigma_p^2/N} - \frac{(X - \langle{x}\rangle)^2}{2\sigma_x^2/N}\right\}}. \label{imDist}
\end{eqnarray}
In the above, ${\cal N}_1$ and ${\cal N}_2$ are normalising factors, and
\begin{eqnarray}
\langle xp \rangle_c &=& \frac{1}{2}\langle \hat x\hat p + \hat p\hat x \rangle - \langle \hat x \rangle\langle\hat p \rangle,\\
\theta_+ &=& \frac{1}{2}\arctan\left(\frac{2\langle x p\rangle_c }{\sigma_x^2-\sigma_p^2}\right),\\
\Delta_+ &=& \sqrt{(\sigma_x^2-\sigma_p^2)^2 + 4\langle x p\rangle_c^2},
\end{eqnarray}
and also
\begin{eqnarray}
\langle xp \rangle_- &=& i\langle \hat x\hat p - \hat p\hat x \rangle,\\
\theta_- &=& \frac{1}{2}\arctan\left(\frac{2\langle x p\rangle_- }{\sigma_x^2-\sigma_p^2}\right),\\
\Delta_- &=& \sqrt{(\sigma_x^2-\sigma_p^2)^2 + 4\langle x p\rangle_-^2}.
\end{eqnarray}

It is noted that the probability distribution for the imaginary parts, ${\cal P}_{im}(X,P)$, explicitly contains the commutator of $\hat x$ and $\hat p$ in the quantities $\theta_-$ and $\Delta_-$. In fact, were $\hat x$ and $\hat p$ commutative then $\theta_- =0$ and
\begin{eqnarray}
{\cal P}_{im}(X,P) &=& 0.
\end{eqnarray}
%This probability distribution encapsulates, in a sense, the `quantumness' of our system.

For the probability distribution for the real parts, ${\cal P}_{re}(X,P)$, we have a product of gaussian distributions mixing combinations of the two generally non-commuting variables $\hat X$ and $\hat P$. However, were $\langle x p\rangle_c = 0$ then we would have a factorisation into two gaussian distributions in $\hat X$ and $\hat P$ separately.

As a special case, upon the substitution
\begin{eqnarray}
\hat P \to \otimes_{i=1}^N \hat {\mathbf 1}_i
\end{eqnarray}
in~(\ref{QCLT}), the probability distribution ${\cal P}_{im}(X,P)$ vanishes and the remaining distribution ${\cal P}_{re}(X,P)$ reduces to a product of distributions of single variable in~(\ref{gaussian}). Alternatively, we could get these same results as with~(\ref{gaussian}) by letting $n=0$ in~(\ref{QCLT}).

\section{Emergence of Classicality}\label{sec5}
From the results of the last section, we can readily derive the following expectation values
\begin{eqnarray}
\left\langle\Phi\left| \hat{X} \right|\Phi\right\rangle
&=& \langle x\rangle;
\end{eqnarray}
and
\begin{eqnarray}
\Sigma_X^2 &\equiv&\left\langle\Phi\left| \hat{X}^2 \right|\Phi\right\rangle -
\left\langle\Phi\left| \hat{X} \right|\Phi\right\rangle^2, \nonumber\\
&\stackrel{N\to \infty}{\sim}& \frac{1}{2N}\left( \sigma_x^2 + \sigma_p^2 +\Delta_+\cos (2\theta_+)\right),\nonumber\\
&\stackrel{N\to \infty}{\sim}& \sigma_x^2/N.
\end{eqnarray}
Similarly,
\begin{eqnarray}
\left\langle\Phi\left| \hat{P} \right|\Phi\right\rangle
&=& \langle p\rangle;
\end{eqnarray}
and
\begin{eqnarray}
\Sigma_P^2 &\equiv&\left\langle\Phi\left| \hat{P}^2 \right|\Phi\right\rangle -
\left\langle\Phi\left| \hat{P} \right|\Phi\right\rangle^2,
\nonumber\\
&\stackrel{N\to \infty}{\sim}& \frac{1}{2N}\left( \sigma_x^2 + \sigma_p^2 -\Delta_+\cos (2\theta_+)\right),\nonumber\\
&\stackrel{N\to \infty}{\sim}&  \sigma_p^2/N.
\end{eqnarray}

Furthermore, it can be shown that the correlation between the coarse-grained/renormalisation block variables $\hat X$ and $\hat P$
\begin{eqnarray}
\frac{1}{2}\left\langle\Phi\left| \hat X\hat P + \hat P\hat X \right|\Phi\right\rangle &
\stackrel{N\to \infty}{\sim}& \int dXdP\; XP \; {\cal P}_{re}(X,P),\nonumber\\
&\stackrel{N\to \infty}{\sim}& 
\langle x\rangle \langle p\rangle + \langle xp \rangle_c/N,\nonumber\\
&\stackrel{N\to \infty}{\sim}& 
\left\langle\Phi\left| \hat{X} \right|\Phi\right\rangle\left\langle\Phi\left| \hat{P} \right|\Phi\right\rangle
+ O(1/N),
\end{eqnarray}
indicating that, in this limit, the coarse-grained/renormalisation block variables are uncorrelated and behaving as classically independent variables. 

For the expectation value of the  {\it hermitian} commutator,
$ i\left\langle\Phi\left| \hat X\hat P - \hat P\hat X \right|\Phi\right\rangle$, we integrate~(\ref{QCLT}) with the distribution ${\cal P}_{im}(X,P)$ for the imaginary part~(\ref{imDist}) to obtain
\begin{eqnarray}
i\left\langle\Phi\left| \hat X\hat P - \hat P\hat X \right|\Phi\right\rangle &\stackrel{N\to \infty}{\sim}& 
\Delta_-\sin(2\theta_-)/2N = \langle xp \rangle_-/N.\label{commutator}
\label{comm}
\end{eqnarray}
%This result is in agreement with~(\ref{classicality}).
It thus follows also that were $\langle \hat x\hat p - \hat p\hat x \rangle=0$
then so would be $\left\langle\Phi\left| \hat X\hat P - \hat P\hat X \right|\Phi\right\rangle=0$, identically for any value of $N$.

We further observe that, in the limit of infinitely many identical and non-interacting quantum subsystems, $N\to\infty$,
\begin{eqnarray}
{\cal P}_{re}(X,P) &\stackrel{N\to\infty}{\longrightarrow}& 
\delta\left((X - \langle{x}\rangle)\cos\theta_+ + (P - \langle{p}\rangle)\sin\theta_+\right)\delta\left((P - \langle{p}\rangle)\cos\theta_+ - (X - \langle{x}\rangle)\sin\theta_+\right),\nonumber\\
&\stackrel{N\to\infty}{\longrightarrow}& \delta\left((X - \langle{x}\rangle)/\cos\theta_+\right)
\delta\left((P - \langle{p}\rangle)\cos\theta_+\right),\nonumber\\
&\stackrel{N\to\infty}{\longrightarrow}& \delta(X - \langle{x}\rangle)\delta(P - \langle{p}\rangle).
\end{eqnarray}

And
\begin{eqnarray}
{\cal P}_{im}(X,P) &\stackrel{N\to\infty}{\longrightarrow}& 
\delta\left((X - \langle{x}\rangle)\cos\theta_- + (P - \langle{p}\rangle)\sin\theta_-\right)\delta\left((P - \langle{p}\rangle)\cos\theta_- - (X - \langle{x}\rangle)\sin\theta_-\right)\nonumber\\
&& -\delta(X - \langle{x}\rangle)\delta(P - \langle{p}\rangle),\nonumber\\
&\stackrel{N\to\infty}{\longrightarrow}& 0.
\end{eqnarray}

Thus,
\begin{eqnarray}
\left\langle\Phi\left|
\sum_{mn} \left(c_{mn} X^m P^n + c^*_{mn} P^n X^m \right)
\right|\Phi\right\rangle &\stackrel{N\to\infty}{\longrightarrow}&
2\sum_{mn} \Re(c_{mn}) \langle x\rangle^m \langle p\rangle^n. \label{classicality}
\end{eqnarray}
The right hand side above now involves only $\langle x\rangle^i$ and $ \langle p\rangle^j$ (with some integers $i$ and $j$), and contains neither $\langle x^i\rangle$ nor $\langle p^j\rangle$, nor the quantum correlations $\langle x^ip^j\rangle$. Implied also in this last expression, which does not include the imaginary parts $\Im(c_{mn})$, is that the expectation value of the commutator of the coarse-grained/renormalisation block variables $\hat X$ and $\hat P$ is vanishingly small with sufficiently large $N$, in agreement with~(\ref{commutator}).

In general, any {\it classical} observable can be expressed indeed as a restricted form of the left hand side of~(\ref{classicality}) with {\it real} $c_{mn}$ -- thus removing the need to consider the distribution for the imaginary part ${\cal P}_{im}(X,P)$.

As a consequence, a regime of classicality could be emerging due to the fact that quantum correlations and all traces of quantum behaviours are now suppressed, except those inherent in the quantum expectation values $\langle x\rangle$ and $\langle p\rangle$. 
\section{Differential Entropies}
A direct generalisation of information Shannon entropy for discrete probabilities $p_d$~\cite{Shannon1}
\begin{eqnarray}
S_d &=& -k_B\sum_i p_d^{(i)}\ln p_d^{(i)}
\end{eqnarray}
to the case of continuous probability distributions might be
\begin{eqnarray}
DEnt_1 &=& -k_B\int Pr(X,P)\ln Pr(X,P)\; dX dP. \label{entropy}
\end{eqnarray}
This is normally called the differential entropy.

This definition of differential entropy, however, does not share all properties of discrete entropy. For example, the differential entropy above can be negative; more importantly, it is not invariant under continuous coordinate transformations. In fact,  Jaynes~\cite{Jaynes} showed that the expression above is not the correct limit of the expression for a finite set of probabilities.

He introduced a modification of differential entropy to address defects in the initial definition of differential entropy by adding an invariant measure factor to correct this~\cite{Jaynes}.

In information theory, this is the limiting density of discrete points in an adjustment to the formula of Shannon for differential entropy.

In the phase space volume $\Delta X_i\Delta P_i$, the transition from discrete probability to continuous probability density should be
\begin{eqnarray}
p_d^{(i)} &\to& Pr(X_i,P_i)\Delta X_i\Delta P_i.
\end{eqnarray}
If this passage to the limit is sufficiently well behaved, we would have
\begin{eqnarray} 
\lim_{N_X\to\infty} \frac{1}{N_X}({\rm number\; of\; points\; in\; [X_i,X_i + \Delta X_i]})
&=& \int_{X_i}^{X_i + \Delta X_i} m(X) dX,
\end{eqnarray}
where $N_X$ is the number of points in the $X$ dimension, and $m(X_i)$ is the density in this dimension. As a result, the differences $\Delta X_i$ in the neighbourhood of any particular value of $X_i$ will have to be
\begin{eqnarray}
\lim_{N_X\to\infty} N_X \Delta X_i = [m(X_i)]^{-1}.
\end{eqnarray}
We have, on the other hand, for the probability density 
\begin{eqnarray}
m(X_i) &=& \int Pr(X_i,P) dP.
\end{eqnarray}
Thus,
\begin{eqnarray}
\lim_{N_X\to\infty} N_X \Delta X_i = \left[\int Pr(X_i,P) dP\right]^{-1}.
\end{eqnarray}
Similarly,
\begin{eqnarray}
\lim_{N_P\to\infty} N_P \Delta P_i = \left[\int Pr(X,P_i) dX\right]^{-1}.
\end{eqnarray}

Putting the above altogether, we have
\begin{eqnarray}
DiffEnt &=& \lim_{N_X,\;N_P\to\infty}\nonumber\\
&& -k_B\sum_i Pr(X_i,P_i) 
\ln\left(\frac{Pr(X_i,P_i)}{N_X N_P \int Pr(X_i,P) dP \int Pr(X,P_i) dX}\right)
\Delta X_i \Delta P_i,\nonumber\\
&=& -k_B\int Pr(X,P) 
\ln\left(\frac{Pr(X,P)}{\int Pr(X,P) dP \int Pr(X,P) dX}\right) dX dP\nonumber\\
&& + \lim_{N_X,\;N_P\to\infty} k_B\ln(N_X N_P).
\end{eqnarray}
From hereon we adopt, following Jaynes, the above as a modified differential entropy, but without the second term, which is infinitely large in the limit, and without the minus sign for the first term to keep our entropy definition semi-positive,
\begin{eqnarray}
DEnt &=& k_B\int Pr(X,P) 
\ln\left(\frac{Pr(X,P)}{\int Pr(X,P) dP \int Pr(X,P) dX}\right) dX dP. \label{diffEntropy}
\end{eqnarray}

This entropy notion is a special instance of the  relative entropy in information theory, also known as the Kullback–Leibler divergence~\cite{Kullback-Leibler} or relative entropy. It is a statistical distance to measure how one probability distribution is different from a second reference probability distribution. A simple interpretation of this divergence is the expected excess surprise from using the latter as a model when the actual distribution is the reference distribution.

We will now investigate the time dependence of such entropy for some systems of non-interacting components. It is noted, and will be illustrated in the next Section, that it is the quantum origin of the non-factorisation of $Pr(X,P)$~(\ref{QCLT}) into component distributions of $X$ and $P$ that gives rise to some interesting and non-trivial temporal behaviours of the entropies.

\section{Time-dependent entropies of some simple systems}\label{sec6}
There are in the literature some considerations of so-called joint entropy for some simple quantum mechanical systems of a single particle~\cite{Dunkel, Garbaczewski}. In this paper in the below we consider, in contrast, certain entropies of composite systems when the number of constituents is infinitely large.

Restricting ourselves to observables in general, it suffices to consider only the particular case whereby $c_{mn}$ in~(\ref{QCLT}) are real. Substituting the probability distribution for the real component~(\ref{reDist}) (which suffices for classical observables) into our adopted entropy~(\ref{diffEntropy}), we arrive at
\begin{eqnarray}
DEnt &=& -k_B\ln\left[(\sigma^2_x\sigma^2_p -\langle x p \rangle_c)/\sigma^2_x\sigma^2_p \right]. \label{diffEntropy2}
\end{eqnarray}
We see from this explicit expression that the non vanishing of $\langle x p \rangle_c$ in general, due to quantum correlations, that enables some non-trivial time dependence for the differential entropy.

\subsection{Free particles}

% Entropy of an ideal gas for three dimensions, according % to Sackur-Tetrode equation,
% \begin{eqnarray}
% S &=& \\
% \Delta S &=& nk\ln \frac{V_f}{V_i}. (\rm isothermal)
% \end{eqnarray}
% On the other hand, from equation~(\ref{entropy})
% \begin{eqnarray}
% \Delta S_P &=& k\ln \frac{\sigma_X(t_f)}{\sigma_X(t_i)}
% \end{eqnarray}

% $n=1$? $n$ is the degree of freedom, for centre of mass it is 1.

For free particles in one dimension, we have for the individual constituent, in the Heisenberg picture,
\begin{eqnarray}
\hat H &=& \frac{\hat p^2}{2m},\nonumber\\
\hat x(t) &=& \hat x(0) + \frac{\hat p(0)}{m}t,\\
\hat p(t) &=& \hat p(0) = \hat p. \nonumber
\end{eqnarray}
The time-dependent variance of the centre of mass, with finite initial variances $\sigma_x^2(0)$ and $\sigma_p^2(0)$, assumes the following temporal behaviours:
\begin{eqnarray}
\sigma_p(t) &=& {\rm constant},
\end{eqnarray}
and
\begin{eqnarray}
\sigma_x^2(t) &=& \left(\sigma_x^2(0) + \frac{t^2}{m^2}\sigma_p^2  + \frac{t}{m}\left(\langle \hat x(0)\hat p + \hat p\hat x(0)\rangle - 2\langle \hat x(0)\rangle\langle \hat p\rangle\right)\right).
%&\stackrel{t\to\infty}{\longrightarrow}& {\cal O}(t^2), \nonumber
\end{eqnarray}

It then follows that the coarse-grained entropy ~(\ref{diffEntropy2}), for a sizable collection of $N$ free and independent particles and for sufficiently large time, is behaving as %${\cal O}(\ln|t|)$,
\begin{eqnarray}
{DEnt}(t) \stackrel{t\to\infty}{\longrightarrow} {\cal O} (\ln|t|).
\end{eqnarray}
which is increasing irreversibly with time (unless the individual subsystem is initially in a momentum eigenstate, whereby $\sigma^2_p(0)= 0 = \langle x(0)p(0)\rangle_c $). Such entropy is increasing with time although invariant with time-reversal, $t \to -t$ and $\hat p \to -\hat p$ -- as is the symmetry of the underlying dynamics of an individual constituent particle.

% Note that the time may have to satisfy the condition~(\ref{limit}) for the CLT to be applicable,
%\begin{eqnarray}
%1 \ll t &\ll& N\frac{\langle \hat p^2(0) \rangle}{\langle \hat p^3(0) \rangle},
%\end{eqnarray}
%which is dependent on the initial wave function of the constituent. (Note that for Gaussian initial momentum wave packet, $\langle \hat p^3(0) \rangle = 0$.)

%Note also that the condition~(\ref{assumption}) is not needed and in fact it is not satisfied in general
%\begin{eqnarray}
%\langle \hat x(t) \hat p(t) \rangle &=& \langle \hat x(0) \hat p(0) + p^2(0)t/m \rangle\nonumber\\
%&\not=& \langle \hat x(0) +\hat p(0)t/m\rangle\langle \hat p(0) \rangle = \langle \hat x(t) \rangle\langle \hat p(t) \rangle.\nonumber
%\end{eqnarray}

\subsection{Uniform and constant force}
For a system under an uniform and constant external force, we have in the Heisenberg picture
\begin{eqnarray}
\hat H &=& \frac{\hat p^2}{2m}-a\hat x,\nonumber\\
\hat x(t) &=& \hat x(0) + {\hat p(0)t/m}+at^2/2m,\\
\hat p(t) &=& \hat p(0)+at. \nonumber
\end{eqnarray}
From which follow the time dependence
\begin{eqnarray}
\sigma_x^2(t)
&=& \left(\sigma_x^2(0) + \frac{t^2}{m^2}\sigma_p(0)^2  + \frac{t}{m}\left(\langle \hat x(0)\hat p(0) + \hat p(0)\hat x(0)\rangle - 2\langle \hat x(0)\rangle\langle \hat p(0)\rangle\right)\right),
%&\stackrel{t\to\infty}{\longrightarrow}& {\cal O}(t^2). \nonumber
\end{eqnarray}
and
\begin{eqnarray}
\sigma^2_p(t) &=& \sigma^2_p(0).
\end{eqnarray}

Upon which, the coarse-grained entropy is, for large time, also increasing irreversibly, 
\begin{eqnarray}
{DEnt} &\stackrel{t\to\infty}{\longrightarrow}&{\cal O}(\ln |t|),
\end{eqnarray}
unless $\sigma^2_p(0)=0$ and $\langle x(0)p(0)\rangle_c = 0$, that is, when the individual subsystem is in a momentum eigenstate initially. Initial position eigenstate is also not applicable here because that would imply an unbounded variance of the momentum due to quantum uncertainty relation.

\subsection{Oscillatory particles}
On the other hand, an example in which the differential entropy is not monotonic in time is that of the quantum simple harmonic oscillator,
\begin{eqnarray}
\hat H &=& \frac{\hat p^2}{2m} + \frac{1}{2}m\omega^2\hat x^2,\nonumber\\
\hat x(t) &=& \hat x(0)\cos(\omega t) + \frac{\hat p(0)}{m\omega}\sin(\omega t),\\
\hat p(t) &=& \hat p(0)\cos(\omega t) - m\omega{\hat x(0)}\sin(\omega t). \nonumber
\end{eqnarray}
From which,
\begin{eqnarray}
\sigma_x^2(t) &=& \cos^2(\omega t)\sigma_x^2(0) + \frac{\sin^2(\omega t)}{m^2\omega^2}\sigma_p^2(0) + 2\frac{\cos(\omega t)\sin(\omega t)}{m\omega}\langle \hat x(0)\hat p(0)\rangle_c,
\end{eqnarray}
and
%which is oscillatory in time. Similarly oscillatory is the momentum variance $\sigma_p(t)$,
\begin{eqnarray}
\sigma_p^2(t) &=& \cos^2(\omega t)\sigma_p^2(0) + {m^2\omega^2}{\sin^2(\omega t)}\sigma_x^2(0) - 2{m\omega}{\cos(\omega t)\sin(\omega t)}\langle \hat x(0)\hat p(0)\rangle_c.
\end{eqnarray}
%and
%\begin{eqnarray}
%\langle \hat x(t)\hat p(t)\rangle_c
%&=& -m\omega\cos(\omega t)\sin(\omega t)\sigma_x^2(0) + \frac{\cos(\omega t)\sin(\omega t)}{m\omega}\sigma_p^2(0) + \nonumber\\
%&& + (\cos^2(\omega t)-\sin^2(\omega t))\langle\hat x(0) \hat p(0)\rangle_c.
%\end{eqnarray}
%Thus
%\begin{eqnarray}
%\sigma_x^2(t)\sigma_p^2(t) - \langle\hat x(t) \hat p(t)\rangle_c^2
%&=& -(4\cos^2(\omega t)\sin^2(\omega t)+(\cos^2(\omega t)-\sin^2(\omega t))^2)\langle\hat x(0) \hat p(0)\rangle_c^2 +\nonumber\\
%&& + (\cos^4(\omega t)+\sin^4(\omega t)+2\cos^2(\omega t)\sin^2(\omega t))\sigma_x^2(0)\sigma_p^2(0) +\nonumber\\
%&& + (-2m\omega\cos(\omega t)\sin(\omega t)(\cos^2(\omega t) -\sin^2(\omega t))+ )\sigma_x^2(0)\langle\hat x(t) \hat p(t)\rangle_c\nonumber\\
%&& + (2\frac{\cos(\omega t)\sin(\omega t)}{m\omega}(\cos^2(\omega t) -\sin^2(\omega t))+)\sigma_p^2(0)\langle\hat x(t) \hat p(t)\rangle_c\nonumber\\
%&=& \sigma_x^2(0)\sigma_p^2(0)-\langle\hat x(0) \hat p(0)\rangle_c^2.
%\end{eqnarray}

In this case, the differential entropy~(\ref{diffEntropy2}) is not, even for large time, a monotonic function of the time.

\section{Summary and concluding remarks}
We derive some quantum mechanical versions of Central Limit Theorems for expectation values of coarse-grained observables, which are functions of coarse-grained hermitean operators. In the above, the coarse-grained variables considered correspond to the center of mass and its classical velocity.

Our derivation methodology could also be rephrased explicitly as a restricted form of renormalisation blocking applied only for observables, and not for non-hermitean operators. Even though incomplete in that sense, the restricted renormalisation is important and useful enough for consideration of all the bulk behaviours that are observable and measurable.

From such hermicity constraints, we obtain for the expectation values positive-definite distributions, which also are the fixed points of the restricted renormalisation group flows. Our probability distributions are also {\it unique} and {\it independent} of the operators being considered. Those are the results in~(\ref{gaussian}) for functions of single macroscopically coarse-grained variables and that in~(\ref{QCLT}) for functions of  macroscopically coarse-grained non-commutative quantum variables. In the latter case, we have two separate distributions for the real and imaginary parts~(\ref{reDist}) and~(\ref{imDist}), respectively -- even though we need only consider the real part for observables. %It is the probability distributions for the imaginary parts that explicitly contain non-trivial quantum commutators and thus encapsulate, in a way, the `quantumness' of our system.

Furthermore, our results herein could be applied also to systems of interacting constituents when approximations whereby the many-body problem could be essentially reduced to a one-body problem, like the mean field Hartree method, are applicable.

Our probability distributions enable a path way for emergence of classical coarse-graining behaviours, as far as observable and measurable, in the limit of an infinitely large number of identical and non-interacting quantum constituents (having finite variances for relevant variables). This is the result of the fact that quantum correlations and all traces of quantum behaviours are now suppressed as shown in~(\ref{classicality}), except those inherent in $\langle x\rangle$ and $\langle p\rangle$ of the constituents.

It should be emphasised that this particular mechanism for such emergence is entirely due to coarse graining in the macroscopic limit, and neither because of environmental decoherence nor due to some kinds of interactions among the constituents.

It is important to note that, because the wave functions are time-dependent in general, in the derivation of the results above we have had to work with a {\it same} time instant for all the microscopic constituent wave functions, as demonstrated by~(\ref{equalTime}). That is, expectation values of the different components $\hat x_i$ and $\hat p_i$ in the block variables $\hat X$ and $\hat P$ must be evaluated at the {\it same} time. This situation is in stark contrast to the classical Central Limit Theorems, which, when dealing with time-independent iid components, can be employed for averaging measurement results over {\it different} moments in time. This distinction is important in our context to recognise that an emergence of classicality would be applicable only for macroscopically block variables -- and not for microscopic variables repeatedly measured and averaged over time. The double-slit experiments could illustrate our point here.  Single electron one by one going through the apparatus still exhibits interference after averaging over many such identical and independent electrons, but macroscopic particles (a macroscopic bunch of many electrons at the same moment of time) may not.

The probability distributions of the quantum Central Limit Theorem further allow us to evaluate some differential entropies for composites of macroscopically coarsed-grained systems. Those entropies are symmetric with respect to time reversal ($t\to -t$, $p\to -p$ and $\phi\to\phi^*$), as is the underlying quantum dynamics. Nevertheless, they could have some interesting and non-trivial temporal dependence. It is noted that it is the quantum origin of the non-factorisation of $Pr(X,P)$~(\ref{QCLT}) into product of component distributions of $X$ and $P$ that gives rise to some interesting and non-trivial temporal behaviours of the entropies. In fact, in some instances, they could also increase with time approximately monotonically -- as functions of the absolute value of the time, for sufficiently large time.

As with the case of classical Central Limit Theorems which have been generalised to cover some less stringent constraints on the behaviours of the constituent components~\cite{Feller}, we expect that further quantum Central Limit Theorems may also be similarly generalised.

%Initial conditions: localisation of an initial drop of ink, or matter-energy at or immediately after the big bang (assuming very weak interactions)!! The Universe as an ideal gas?

%Poincare cycle doesnot apply for the free particles but does for the SHO!

%Coarse-grained
%How large is $N$ is dictated by various single-particle expectation values
%statistical entropy versus Boltzmann
%Stochastic distribution for single-particle dynamics, and all particles are of the same kind (Planck)

%number density

%Coarse-graining <- see preprint "Lecture 6: Entropy" in the same folder

%STATISTCAL FORCE!!

%The same results for classical context -> coarse- graining (reduced) variables and statistical force are responsible for observed irreversibility 

%A sound of thunder
%Arrival
%Tenet

%See Appendix~\ref{app1}

\section*{Acknowledgement}
I want to thank Peter Hannaford for some input for this paper.

\appendix
\section{Derivation of a Quantum Central Limit Theorem for function of a coarse-grained/renormalisation block variable}\label{app2}
We derive the result~(\ref{gaussian}) in this Appendix. 

Consider a system with identical and non-interacting components (like in the case of an ideal gas)
\[|\Phi\rangle = \otimes_i^N |\phi_i\rangle,\]
where $|\phi_i\rangle = |\phi\rangle$, for all $i$.

Now with some function $f$, we consider
\begin{eqnarray}
\cal E &=& \left\langle\Phi \left| f\left(\frac{1}{N}\sum_i^N \hat x_i\right)\right|\Phi \right\rangle
\end{eqnarray}

Insert the resolution of identity into the above
\[\hat 1 = \int \prod_i^N dx_i |x_i \rangle\langle x_i|,\]
where
\begin{eqnarray}
\hat x_i |y_i\rangle &=& y_i|y_i\rangle. \nonumber
\end{eqnarray}
We then have
\begin{eqnarray}
\cal E &=& \left\langle\Phi \left| \left( \int \prod_j^N dy_j |y_j \rangle\langle y_i| \right)\int  f\left(\frac{1}{N}\sum_i^N \hat x_i\right) \prod_k^N dx_k|x_k \rangle\langle x_k|\right|\Phi \right\rangle \nonumber\\
&=& \int  f\left(\frac{1}{N}\sum_i^N x_i\right) \prod_k^N |\langle x_k|\phi_k \rangle|^2 dx_k
\end{eqnarray}
Insert the identity
\[ 1 = \int dX\; \delta\left( X - \frac{1}{N}\sum_j^N x_j\right),\]
in which the delta function can be expressed as
\[\delta (u) = \frac{1}{2\pi}\int dw \;{\rm e}^{iwu}. \]
We further obtain
\begin{eqnarray}
\cal E &=& \int dX \int \delta\left( X - \frac{1}{N}\sum_j^N x_j\right) f\left(\frac{1}{N}\sum_k^N x_k\right) \prod_i^N |\langle x_i|\phi_i \rangle|^2 dx_i, \nonumber\\
&=& \frac{1}{2\pi}\int dX\;dw\;f(X)\; {\rm e}^{iXw}\left[\int dx\;{\rm e}^{-ixw/N} |\langle x|\phi \rangle|^2\right]^N,\label{equalTime}
\end{eqnarray}
It should be emphasised that the probability distributions of the constituents $|\langle x|\phi_i \rangle|^2$ are functions of the time in general; and that in arriving at~(\ref{equalTime}) we have had to take the {\it same} instant of time for all the component probabilities.
\begin{eqnarray}
\cal E &=& \frac{1}{2\pi}\int dX\;dw\;f(X)\; {\rm e}^{iXw}\left[\int dx\;(1-iwx/N - w^2x^2/2N^2 + O(1/N^3))|\langle x|\phi \rangle|^2\right]^N,\nonumber\\
&=& \frac{1}{2\pi}\int dX\;dw\;f(X)\; {\rm e}^{iXw}\left[\;1-iw\langle{x}\rangle/N - w^2\langle{x^2}\rangle/2N^2 + O(1/N^3)\right]^N,\nonumber\\
&=& \frac{1}{2\pi}\int dX\;dw\;f(X)\; {\rm e}^{iXw}\;\exp \{-iw\langle{x}\rangle - w^2(\langle{x^2}\rangle-\langle{x}\rangle^2)/2N + O(1/N^2)\}.\label{intermediate}
% &=& \frac{1}{2\pi}\int dX\; f(X)\; \int dw\; {\rm e}^{iw(X-\langle{x}\rangle) - w^2(\langle{x^2}\rangle-\langle{x}\rangle^2)/2N + O(1/N^2) } \}.\label{intermediate}
\end{eqnarray}

Integrating over $w$, we finally arrive at the result
\begin{eqnarray}
\left\langle\Phi \left| f\left(\frac{1}{N}\sum_i^N \hat x_i\right)\right|\Phi \right\rangle 
&\stackrel{N\to \infty}{\longrightarrow}& \frac{1}{(\sigma_x/\sqrt N)2\pi \sqrt{2\pi}}\int dX f(X) \exp{\left\{-\frac{(X - \langle{x}\rangle)^2}{2(\sigma_x/\sqrt N)^2}\right\}},\label{A1}
\end{eqnarray}
where in (\ref{intermediate}) and (\ref{A1}), we have defined
\begin{eqnarray}
\langle{x}\rangle &\equiv& \int x |\langle x|\phi \rangle|^2 dx  = \langle \phi|\hat x|\phi\rangle,\\
\sigma_x^2 &\equiv& \langle{x^2}\rangle - \langle{x}\rangle^2.
\end{eqnarray}

\section{Derivation of a Quantum Central Limit Theorem for non-commuting operators}\label{app1}
In this Appendix we derive the result~(\ref{QCLT}) for our Quantum Central Limit Theorem. 

Let us first consider the expectation value,
\begin{eqnarray}
{\cal G} &=& \left\langle\Phi \left| (\hat X)^m (\hat P)^n + (\hat P)^n (\hat X)^m\right|\Phi \right\rangle,\nonumber\\
&=& \left\langle
\Phi \left| \left(\int \prod_i^N dy_i |y_i \rangle\langle y_i|\right)(\hat X)^m (\hat P)^n\left(\int \prod_i^N dk_i |k_i \rangle\langle k_i|\right)\right.\right. +\nonumber\\
&& \left.\left.+\left(\int \prod_i^N dk_i |k_i \rangle\langle k_i|\right)(\hat P)^n (\hat X)^m\left(\int \prod_i^N dy_i |y_i \rangle\langle y_i|\right)\right|\Phi\right\rangle,\nonumber\\
&=& \int  \left(\sum_i^N y_i/N\right)^m\left(\sum_i^N k_i/N\right)^n \nonumber\\
&&\left(\prod_{i}^N \langle\phi_i |y_i\rangle\langle y_i|k_i \rangle\langle k_i|\phi_i\rangle dy_i dk_i + \prod_{i}^N \langle\phi_i |k_i\rangle\langle k_i|y_i \rangle\langle y_i|\phi_i \rangle dy_i dk_i\right), \label{19}
\end{eqnarray}
where we have inserted the resolutions of identity, respectively, for the $x$ and $p$-representation,
\begin{eqnarray}
\hat 1 = \int \prod_i^N dy_i |y_i \rangle\langle y_i|,\\
\hat 1 = \int \prod_i^N dk_i |k_i \rangle\langle k_i|,
\end{eqnarray}
in which the eigenvectors of position and momentum, respectively, satisfy
\begin{eqnarray}
\hat x_i |y_i\rangle &=& y_i|y_i\rangle,\\
\hat p_i |k_i\rangle &=& k_i|k_i\rangle.
\end{eqnarray}

We next insert to the integrand of~(\ref{19}) the identities
\begin{eqnarray}
1 = \int dX\; \delta\left( X - \frac{1}{N}\sum_j^N y_j\right) =  \int dP\; \delta\left( P - \frac{1}{N}\sum_j^N k_j\right),
\end{eqnarray}
in which the delta functions can also be expressed as
\begin{eqnarray}
\delta \left( X - \sum_j^N y_j/N\right) = \frac{1}{2\pi}\int dw \;{\rm e}^{iw( X - \sum_j^N y_j/N)},\\
\delta \left( P - \sum_j^N p_j/N\right) = \frac{1}{2\pi}\int d\lambda \;{\rm e}^{i\lambda( P - \sum_j^N p_j/N)}.
\end{eqnarray}

From~(\ref{19}) we then have
\begin{eqnarray}
{\cal G} &=& \frac{1}{2}\int dX dP\;\delta\left( X - \frac{1}{N}\sum_j^N y_j\right) \delta\left( P - \frac{1}{N}\sum_j^N k_j\right) g\left(\frac{1}{N}\sum_i^N y_i, \frac{1}{N}\sum_i^N k_i\right) \nonumber \\
&&\left( \prod_{i}^N \langle\phi_i |y_i\rangle\langle y_i|k_i \rangle\langle k_i|\phi_i\rangle dy_i dk_i + \prod_{i}^N \langle\phi_i |k_i\rangle\langle k_i|y_i \rangle\langle y_i|\phi_i \rangle dy_i dk_i\right), \nonumber\\
&=& \frac{1}{8\pi^2}\int dXdP\;dwd\lambda\;g(X,P)\; {\rm e}^{iXw+iP\lambda}\nonumber\\
&&\left[\left(\int dydk\;{\rm e}^{-iyw/N-ik\lambda/N} \langle\phi |y\rangle\langle y|k \rangle\langle k|\phi \rangle \right)^N + \left(\int dydk\;{\rm e}^{-iyw/N-ik\lambda/N}\langle\phi |k\rangle\langle k|y \rangle\langle y|\phi \rangle\right)^N\right],\nonumber\\
&=& \frac{1}{8\pi^2}\int dXdP\;dwd\lambda\;g(X,P)\; {\rm e}^{iXw+iP\lambda}\nonumber\\
&&\left[\left(\int dydk\;\left(1-i(wy+\lambda k)/N - (wy+\lambda k)^2/2N^2 + O(1/N^3)\right)\langle\phi |y\rangle\langle y|k \rangle\langle k|\phi \rangle\right)^N\right.\nonumber\\
&& + \left.
\left(
\int dydk\;\left(1-i(wy+\lambda k)/N - (wy+\lambda k)^2/2N^2 + O(1/N^3)\right)
\langle\phi |k\rangle\langle k|y \rangle\langle y|\phi \rangle\right)^N
\right],\nonumber\\
&=& \frac{1}{8\pi^2}\int dXdP\;dwd\lambda\;g(X,P)\; {\rm e}^{iXw+iP\lambda}\nonumber\\
&&\left[\;\left(1-iw\overline{y}/N - i\lambda\overline{k}/N -\overline{(wy+\lambda k)^2}/2N^2 + O(1/N^3)\right)^N +\right.\nonumber\\
&& \left.
+ \left(1-iw\widetilde{y}/N - i\lambda\widetilde{k}/N -\widetilde{(wy+\lambda k)^2}/2N^2 + O(1/N^3)\right)^N\right],\nonumber\\
&=& \frac{1}{8\pi^2}\int dXdP\;dwd\lambda\;g(X,P)\; {\rm e}^{iXw+iP\lambda}\nonumber\\
&& \left[\exp \left\{-iw\overline{y} -i\lambda\overline{k} - (\overline{(wy +\lambda k)^2}-(w\overline{y} +\lambda \overline{k})^2)/2N + O(1/N^2)\right\}+\right.,\nonumber\\
&& \left. + \exp \left\{-iw\widetilde{y} -i\lambda\widetilde{k} - (\widetilde{(wy +\lambda k)^2}-(w\widetilde{y} +\lambda \widetilde{k})^2)/2N + O(1/N^2)\right\}\right].
\label{intermediate2}
% &=& \frac{1}{2\pi}\int dX\; f(X)\; \int dw\; {\rm e}^{iw(X-\langle{x}\rangle) - w^2(\langle{x^2}\rangle-\langle{x}\rangle^2)/2N + O(1/N^2) } \}.\label{intermediate}
\end{eqnarray}
In the last two expressions we have introduced the notations
\begin{eqnarray}
\overline {c(y,k)} &\equiv& \int dydk\; c(y,k)\; \langle\phi |y\rangle\langle y|k \rangle\langle k|\phi \rangle,\\
\widetilde {c(y,k)} &\equiv& \int dydk\; c(y,k)\; \langle\phi |k\rangle\langle k|y \rangle\langle y|\phi \rangle.
\end{eqnarray}
It can be seen that for any function $a(y)$ of the position $y$, using the completeness of the momentum basis,
\begin{eqnarray}
\overline {a(y)} &=& \int dydk\; a(y)\; \langle\phi |y\rangle\langle y|k \rangle\langle k|\phi \rangle,\nonumber\\
&=& \int dy\; a(y)\; \langle\phi |y\rangle\langle y|\left(\int dk |k \rangle\langle k|\right)|\phi \rangle,\nonumber\\
&=& \int dy\; a(y)\; \langle\phi |y\rangle\langle y|\phi \rangle,\nonumber\\
\overline {a(y)} &=& \langle a(\hat y) \rangle.
\end{eqnarray}
Similarly,
\begin{eqnarray}
\widetilde {a(y)} &=& \int dydk\; a(y)\; \langle\phi |k\rangle\langle k|y \rangle\langle y|\phi \rangle,\nonumber\\
&=& \int dy\; a(y)\; \langle\phi |\left(\int dk |k \rangle\langle k|\right)|y\rangle\langle y|\phi \rangle,\nonumber\\
&=& \int dy\; a(y)\; \langle\phi |y\rangle\langle y|\phi \rangle,\nonumber\\
\widetilde {a(y)} &=& \langle a(\hat y) \rangle.
\end{eqnarray}
Thus,
\begin{eqnarray}
\overline {a(y)} &=& \widetilde {a(y)}  = \langle a(\hat y) \rangle.
\end{eqnarray}
Also for any function $b(k)$,
\begin{eqnarray}
\overline {b(k)} &=& \widetilde {b(k)}  = \langle b(\hat k) \rangle.
\end{eqnarray}

For product of $a(\hat y)b(\hat k)$, 
\begin{eqnarray}
\langle a(\hat y)b(\hat k) \rangle &=& \langle \phi| a(\hat y)b(\hat k) |\phi\rangle,\nonumber\\
&=& \langle \phi| \left(\int dy |y \rangle\langle y|\right)a(\hat y)b(\hat k) \left(\int dk |k \rangle\langle k|\right)|\phi\rangle,\nonumber\\
&=& \int dydk\; a(y)b(k)\; \langle \phi|y \rangle\langle y|k \rangle\langle k|\phi\rangle,\nonumber
\end{eqnarray}
that is,
\begin{eqnarray}
\overline{a(y)b(k)} &=& \langle a(\hat y)b(\hat k) \rangle.
\end{eqnarray}
On the other hand, for the reverse ordering, it also follows that
\begin{eqnarray}
\widetilde{a(y)b(k)} &=&
\langle b(\hat k)a(\hat y) \rangle .
\end{eqnarray}

We now introduce the notation
\begin{eqnarray}
\langle xp \rangle_c &=& \frac{1}{2}\langle \hat x\hat p + \hat p\hat x \rangle - \langle \hat x \rangle\langle\hat p \rangle.
\end{eqnarray}
%%%%%%%%%%%%%%%%%%%
Back to (\ref{intermediate2}), we can rewrite that expression as
\begin{eqnarray}
{\cal G} &\sim& \frac{1}{2}\int dXdP\;dwd\lambda\;g(X,P)\; {\rm e}^{iXw+iP\lambda-iw\langle y\rangle -i\lambda\langle k\rangle}\nonumber\\
&& \left[\exp \left\{ - (w^2\sigma_y^2 +\lambda^2\sigma_k^2 +2w\lambda(\langle \hat y\hat k\rangle - \langle \hat y\rangle\langle\hat k\rangle ))/2N + O(1/N^2)\right\}+\right.\nonumber\\
&& \left. + \exp \left\{ - (w^2\sigma_y^2 +\lambda^2\sigma_k^2 +2w\lambda(\langle \hat k\hat y\rangle - \langle \hat y\rangle\langle\hat k\rangle ))/2N + O(1/N^2)\right\}\right].\nonumber\\
&\sim& \int dXdP\;dwd\lambda\;g(X,P)\; {\rm e}^{iXw+iP\lambda-iw\langle y\rangle -i\lambda\langle k\rangle}\nonumber\\
&& \exp \left\{-(w^2\sigma_y^2 +\lambda^2\sigma_k^2 +2w\lambda\langle y k\rangle_c)/2N + O(1/N^2)\right\}.
\end{eqnarray}

%\fbox{\begin{minipage}{50 em}
Integrating the last expression over $w$ and $\lambda$, we finally obtain, for $N\gg1$,
\begin{eqnarray}
\left\langle\Phi \left| (\hat X)^m (\hat P)^n + (\hat P)^n (\hat X)^m\right|\Phi \right\rangle \nonumber\\
\stackrel{N\to \infty}{\sim}&& 2\int dXdP\; X^mP^n\nonumber\\
&& \times\exp{\left\{-\frac{\left[(X - \langle{x}\rangle)\cos\theta_+ + (P - \langle{p}\rangle)\sin\theta_+\right]^2}{(\sigma_x^2 + \sigma_p^2+\Delta_+)/N}\right\}}\nonumber\\
&&\times\exp{\left\{-\frac{\left[(P - \langle{p}\rangle)\cos\theta_+ - (X - \langle{x}\rangle)\sin\theta_+\right]^2}{(\sigma_x^2 + \sigma_p^2-\Delta_+)/N}\right\}}, \label{gaussian3}\nonumber\\
\end{eqnarray}
where
\begin{eqnarray}
\langle xp \rangle_c &=& \frac{1}{2}\langle \hat x\hat p + \hat p\hat x \rangle - \langle \hat x \rangle\langle\hat p \rangle,\\
\theta_+ &=& \frac{1}{2}\arctan\left(\frac{2\langle x p\rangle_c }{\sigma_x^2-\sigma_p^2}\right),\\
\Delta_+ &=& \sqrt{(\sigma_x^2-\sigma_p^2)^2 + 4\langle x p\rangle_c^2}.
\end{eqnarray}
This is the probability distribution for the real part ${\cal P}_{re}(X,P)$ of~(\ref{reDist}).

Similar to the derivation above, it can also be shown that the distribution for the imaginary part ${\cal P}_{im}(X,P)$ of~(\ref{imDist}) is, for some integers $m$ and $n$,
\begin{eqnarray}
i\left\langle\Phi\left| \hat X^m\hat P^n - \hat P^n\hat X^m \right|\Phi\right\rangle &
\stackrel{N\to \infty}{\sim}& \int dXdP\; X^m P^n\nonumber\\
&& \left({\cal N}_1\exp{\left\{-\frac{\left[(X - \langle{x}\rangle)\cos\theta_- + (P - \langle{p}\rangle)\sin\theta_-\right]^2}{(\sigma_x^2 + \sigma_p^2+\Delta_-)/N}\right\}}\right.\nonumber\\
&&\;\;\;\;\;\times\exp{\left\{-\frac{\left[(P - \langle{p}\rangle)\cos\theta_- - (X - \langle{x}\rangle)\sin\theta_-\right]^2}{(\sigma_x^2 + \sigma_p^2-\Delta_-)/N}\right\}}\nonumber\\
&&-\left.{\cal N}_2\exp{\left\{-\frac{(P - \langle{p}\rangle)^2}{2\sigma_p^2/N} - \frac{(X - \langle{x}\rangle)^2}{2\sigma_x^2/N}\right\}}\right), %\label{gaussian3}
\end{eqnarray}
where
\begin{eqnarray}
\langle xp \rangle_- &=& i\langle \hat x\hat p - \hat p\hat x \rangle,\\
\theta_- &=& \frac{1}{2}\arctan\left(\frac{2\langle x p\rangle_- }{\sigma_x^2-\sigma_p^2}\right),\\
\Delta_- &=& \sqrt{(\sigma_x^2-\sigma_p^2)^2 + 4\langle x p\rangle_-^2}.
\end{eqnarray}

We could recover the results from~(\ref{gaussian}) by either putting $n=0$ or replacing $\hat P$ by $\bf\hat 1$ in~~(\ref{QCLT}).

\bibliography{Ref2}

\end{document}